%% file: AnonymousSubmission2027.tex
\documentclass[letterpaper]{article} 
\usepackage[preprint]{aaai2027}  
\usepackage[hyphens]{url}  
\usepackage{graphicx} 
\usepackage{natbib}  
\usepackage{caption} 
\usepackage{amsmath}
\usepackage{amssymb}
\usepackage{algorithm}
\usepackage{algorithmic}
\usepackage{newfloat}
\usepackage{listings}
\DeclareCaptionStyle{ruled}{labelfont=normalfont,labelsep=colon,strut=off} 
\floatstyle{ruled}
\newfloat{listing}{tb}{lst}{}
\floatname{listing}{Listing}
\usepackage{booktabs}
\usepackage{enumitem}
\usepackage{tikz}
\usetikzlibrary{arrows.meta,positioning,calc,fit,backgrounds}

\title{EduClaw-Bench: A Long-Horizon Benchmark for Pedagogical
LLM Agents with Simulated Learners}

\author{
    Unggi Lee\textsuperscript{\rm 1},
    Sookbun Lee\textsuperscript{\rm 2}\thanks{Corresponding authors.},
    Yeil Jeong\textsuperscript{\rm 3}\footnotemark[1],
    Eunjoo Lee\textsuperscript{\rm 4}\footnotemark[1],
    Minchul Shin\textsuperscript{\rm 4}\footnotemark[1],
    Hoilym Kwon\textsuperscript{\rm 5}
}
\affiliations{
    \textsuperscript{\rm 1}Korea University Sejong Campus,
    \textsuperscript{\rm 2}Opentutorials,
    \textsuperscript{\rm 3}Indiana University,
    \textsuperscript{\rm 4}Gyeonggi Institute of Education,
    \textsuperscript{\rm 5}Korea University\\
    First author: codingchild@korea.ac.kr
}

\begin{document}

\maketitle

\begin{abstract}
Large language models (LLMs) power educational applications from tutoring to essay scoring, but each is a point solution to a single task, and only recently have these point solutions been integrated into agents operating over a learning management system (LMS). Yet tutoring is long-horizon, since a learner improves over days and weeks rather than in a single turn, and no benchmark evaluates an agent tutor across a sustained relationship. We introduce \emph{EduClaw-Bench}, a benchmark that places an agent tutor in a continuous 30-day relationship with a simulated learner grounded in knowledge tracing (KT), whose knowledge-concept mastery, from a KT model trained on real-student data, drives its answers and is probed for learning gain across 55 scenarios. Each agent is scored on three primary axes (learning gain, responsiveness, and helpfulness) and two curriculum-design axes (Gagn\'e and Rosenshine), with helpfulness and the curriculum axes judged by a cross-family panel of three LLM judges. Evaluating 10 agent adapters over three base-model tiers yields two findings that single-tier, single-session evaluation cannot reach. First, tutoring quality belongs to the base model and the agent harness together rather than either alone. Second, almost no combination sustains good tutoring over the full horizon. A calibration check ($\text{ECE}=0.049$) and a live-classroom field study confirm that the simulated learner and its measurements track reality. Our work is a step toward trustworthy AI tutors for future education.
\end{abstract}

\input{sections/01_introduction.tex}

\input{sections/02_related_work.tex}
\input{sections/04_framework.tex}
\input{sections/05_setup.tex}
\input{sections/06_results.tex}

\input{sections/07_analysis.tex}
\input{sections/09_discussion.tex}


\bibliography{aaai2027}

\clearpage
\input{sections/10_appendix.tex}

\end{document}

%% file: sections/01_introduction.tex

\section{Introduction}
\label{sec:intro}

\begin{figure*}[!t]
\centering
\includegraphics[width=\textwidth]{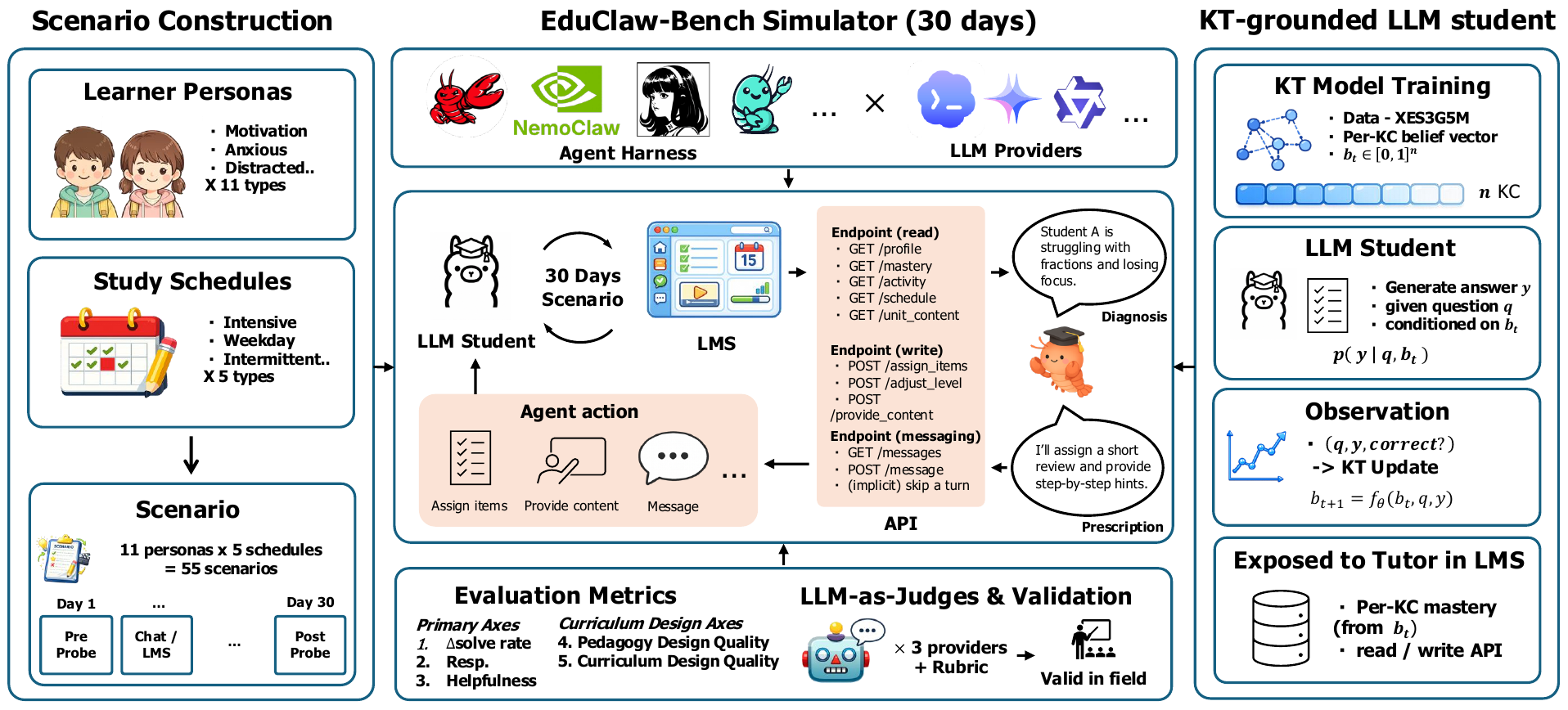}
\caption{Overview of \emph{EduClaw-Bench}. \emph{Left}, each of 55 scenarios pairs one of 11 learner personalities with one of 5 study schedules and runs from a pre-probe on day 1 to a post-probe on day 30. \emph{Center}, an agent tutor on a chosen LLM provider interacts with the student over 30 days through an LMS read, write, and messaging API, and its turns are scored on three primary axes ($\Delta$ Solve Rate, Responsiveness, Helpfulness) plus two curriculum-design axes (Gagn\'e, Rosenshine), the helpfulness and curriculum axes judged by a cross-family panel of three LLM judges. \emph{Right}, the student is grounded by a KT model trained on XES3G5M whose per-KC belief $\mathbf{b}$ conditions each answer and is updated from every observation, and that belief is exposed to the tutor as per-KC mastery through the LMS.}
\label{fig:architecture}
\end{figure*}

Large language models (LLMs) power educational applications, from tutoring and programming feedback to essay scoring~\cite{wang_2024_tutorcopilot,kazemitabaar_2024_codeaid,liffiton_2023_codehelp,cai_2025_essayscoring}, but each targets a single task in isolation. Only recently has a line of work begun to integrate these capabilities into an agent that operates through a learning management system (LMS), assigning work, reading a learner's mastery, and sustaining day-to-day conversation rather than answering one-off questions~\cite{deeptutor_2026}. This shift matters because learning is a long-horizon process. A learner improves over days and weeks of curriculum-structured instruction rather than in a single exchange~\cite{gagne_1985_conditions,rosenshine_2012_principles}, so a tutor's learning gain, curriculum, and durable helpfulness show only over a sustained relationship, not in a single turn~\cite{li_2026_longtutor}.

Existing tutor evaluations do not capture such long-horizon learning, scoring the quality of a turn or session against fixed learner utterances rather than the learning a tutor causes, whether through pedagogical-ability taxonomies~\cite{maurya_2024_tutoreval}, open-ended teaching capabilities~\cite{macina_2025_mathtutorbench}, dialogue datasets~\cite{macina_2023_mathdial}, or interactive math evaluation~\cite{collins_2023_checkmate}. Even the benchmarks nearest to a longitudinal or agentic setting stop short, as LongTutor~\cite{li_2026_longtutor} scores plain LLM tutors on static long-term logs without direct outcomes and EduAgentBench~\cite{eduagentbench_2026} scores agent readiness on teacher-workflow tasks rather than student learning, so none couples an agent tutor to a learner that actually learns. Measuring learning itself needs a simulated learner that persists across days, updates its knowledge from what the tutor does, and can be probed for gain. The natural substitute, an LLM prompted to role-play a student, is exactly where recent work reports failures, as prompted or fine-tuned LLM students do not reproduce real students' ability distributions~\cite{srivatsa2025reliably}, misconception dynamics~\cite{do2026misconception}, or authenticity under tutoring~\cite{scarlatos2026substance}.

We introduce \emph{EduClaw-Bench}\footnote{\url{https://anonymous.4open.science/r/educlaw-bench-anonymous-ED3F}}, a benchmark that puts an agent tutor into a continuous 30-day relationship with a simulated learner grounded in knowledge tracing (KT). Its mastery over knowledge concepts (KCs), estimated by a KT model trained on real students' interaction data, both drives its free-response answers and is probed daily for learning gain. Across 55 scenarios that pair 11 learner personalities with 5 study schedules, each a continuous span of 30 virtual days of chat and LMS interaction, we score a tutor on three primary axes (learning gain, responsiveness, and helpfulness) and two curriculum-design axes (Gagn\'e and Rosenshine). \emph{EduClaw-Bench} evaluates pedagogical agents over a long horizon with KT-grounded simulated learners, laying a foundation for trustworthy AI tutors in future education.

\subsection*{Contributions}
\begin{itemize}[leftmargin=1.2em, itemsep=2pt, topsep=2pt]
\item \emph{EduClaw-Bench}, the first benchmark that puts an agent tutor in a 30-day, LMS-based relationship with a KT-grounded simulated learner, scoring learning gain, responsiveness, helpfulness, and curriculum across 55 persona-and-schedule scenarios.
\item Tutoring quality belongs to the base model and agent harness together rather than either alone, so rankings reorder across tiers and single-tier leaderboards mislead.
\item Almost no model-and-harness combination sustains tutoring over the full horizon, as learning plateaus within days, so long-horizon competence needs a purpose-matched combination.
\end{itemize}

%% file: sections/02_related_work.tex

\section{Related Work}
\label{sec:related}

\paragraph{Benchmarks for pedagogical tutors.}
Most tutor benchmarks score the quality of a tutor's turn or session against fixed learner utterances. Rubric-based evaluations judge pedagogical moves through taxonomies \cite{maurya_2024_tutoreval}, open-ended teaching capabilities \cite{macina_2025_mathtutorbench}, dialogue datasets \cite{macina_2023_mathdial}, interactive math \cite{collins_2023_checkmate}, or learning-science rubrics \cite{learnlm_2025_v2}, and a parallel line trains tutors toward pedagogical objectives \cite{lee_2026_pedagogicalrl_thinking,puerto_2025_pedagogicalrl,sonkar_2024_pedagogical_alignment,scarlatos_2025_trainingtutors,wang_2024_tutorcopilot,kargupta_2024_instruct_not_assist}. Because these measure tutor-side behavior, they need a stand-in learner, yet recent work shows LLM-simulated students do not faithfully reproduce real students' abilities \cite{srivatsa2025reliably}, misconceptions \cite{do2026misconception}, or authenticity under tutoring \cite{scarlatos2026substance}, leaving open how to obtain a learner faithful enough to measure the learning a tutor causes.

\paragraph{Agent tutors over longer horizons.}
Closer to our setting are agent- and long-horizon-oriented benchmarks. LongTutor \cite{li_2026_longtutor} scores plain LLM tutors on static long-term logs without direct outcomes, EduAgentBench \cite{eduagentbench_2026} and ISD-Agent-Bench \cite{jeon2026isdbench} score agent readiness on teacher-side workflow and instructional-design tasks rather than student learning, and DeepTutor's TutorBench \cite{deeptutor_2026} uses source-grounded learner profiles in single sessions. None couples an agent tutor to a learner that actually learns. As Table~\ref{tab:benchmark-compare} summarizes, \emph{EduClaw-Bench} is the only benchmark that pairs agent tutors with a KT-grounded simulated learner to score direct student outcomes on independent axes over 30-day continuous scenarios across multiple tutor base-model tiers.

\begin{table*}[t]
\centering
\scriptsize
\setlength{\tabcolsep}{3pt}
\caption{Comparison of pedagogical LLM tutor benchmarks by learner (R = real logs, S = KT-plus-LLM simulation), agent support, outcome versus proxy scoring, curriculum-design scoring, interaction horizon, and cross-tier comparison. \emph{EduClaw-Bench} is the only benchmark that pairs an agent tutor with a KT-grounded simulated learner and scores direct learning outcomes over a 30-day horizon across multiple base-model tiers, on five axes (three outcome plus two curriculum-design).}
\label{tab:benchmark-compare}
\begin{tabular}{llllllll}
\toprule
Benchmark & Year & Learner & Agent & Outcome scoring & Curriculum & Horizon & Cross-tier \\
\midrule
LearnLM~\cite{learnlm_2025_v2}                    & 2025 & S (human eval)        & no  & rubric only         & no  & session & Gemini variants \\
TutorEval~\cite{maurya_2024_tutoreval}            & 2024 & R                     & no  & rubric only         & no  & turn    & no \\
MathTutorBench~\cite{macina_2025_mathtutorbench}  & 2025 & R                     & no  & rubric only         & no  & turn    & no \\
MathDial~\cite{macina_2023_mathdial}              & 2023 & R                     & no  & dialogue quality    & no  & session & no \\
CheckMate~\cite{collins_2023_checkmate}           & 2023 & R                     & no  & pref.\ vs correctness & no & turn  & no \\
LongTutor~\cite{li_2026_longtutor}                & 2026 & R (static logs)       & no  & 3 formative tasks   & no  & turn    & no \\
EduAgentBench~\cite{eduagentbench_2026}           & 2026 & R (workflows)         & yes & teacher-workflow    & no  & session & no \\
ISD-Agent-Bench~\cite{jeon2026isdbench}           & 2026 & none                  & yes & ISD design tasks    & yes & session & 3 LLMs \\
TutorBench~\cite{deeptutor_2026}                  & 2026 & S (source-grounded)   & yes & personalization gain & no & session & partial \\
\textbf{EduClaw-Bench (ours)}                     & \textbf{2026} & \textbf{S (KT + LLM)} & \textbf{yes} & \textbf{direct (3 axes)} & \textbf{yes (2 axes)} & \textbf{30 days} & \textbf{yes (3 tiers)} \\
\bottomrule
\end{tabular}
\end{table*}

%% file: sections/04_framework.tex

\section{The EduClaw-Bench Framework}
\label{sec:framework}

We built \emph{EduClaw-Bench} as a virtual environment for evaluating pedagogical agents, sketched end to end in Figure~\ref{fig:architecture}. It follows the design pattern established by prior agent-evaluation environments that expose a fixed action space and grade autonomous behavior on sequential tasks (OSWorld \cite{xie_2024_osworld}, WebArena \cite{zhou_2023_webarena}, AgentBench \cite{liu_2023_agentbench}, and $\tau$-bench \cite{yao_2024_taubench}), and adds one dimension those environments do not evaluate, namely long-horizon behavior across a 30-day teacher-learner relationship rather than a single job or session.

\subsection{Virtual Environment}
\label{sec:env}

Each scenario is a 30-day timeline pairing one of 11 learner personalities (10 core such as confused productive, silent struggler, and frustrated spiral, plus one adversarial student used for safety probes) with one of 5 study schedules (after-school, homework-only, school-integrated, weekend-warrior, self-directed). This gives $11 \times 5 = 55$ scenarios. Each scenario file (YAML) fixes daily blocks, LMS event triggers, and probe schedules.

The tutor interacts with the learner through two APIs, summarized in Table~\ref{tab:api}. The messaging API is symmetric (\texttt{GET /messages}, \texttt{POST /message}) and the LMS API is split into five read endpoints (learner state) and three write endpoints (interventions). We distinguish two categories of endpoint call, namely \emph{queries} that observe learner state without modifying it (all \texttt{GET}s), and \emph{actions} that either send a chat turn to the learner or write a change into the LMS. Only the second category is counted as a tutor action, since queries are prerequisites the tutor uses to know what to do, not interventions in their own right. This yields exactly five observable tutor action kinds at the trace level, namely \texttt{send\_message}, \texttt{stay\_silent}, \texttt{assign\_items}, \texttt{adjust\_level}, and \texttt{provide\_content}. Diagnosis (choosing which KCs to target) is a query step through \texttt{GET /mastery}, while intervention is an action step through the three LMS write endpoints.

\begin{table}[t]
\centering
\scriptsize
\setlength{\tabcolsep}{3pt}
\renewcommand{\arraystretch}{1.05}
\caption{The tutor-facing API surface, comprising five LMS read endpoints, three LMS write endpoints, and a messaging channel, with the five observable tutor action kinds in the rightmost column.}
\label{tab:api}
\begin{tabular}{@{}lll@{}}
\toprule
Endpoint & Purpose & Action kind \\
\midrule
\multicolumn{3}{@{}l@{}}{\emph{LMS read}} \\
\texttt{GET /profile}       & learner metadata          & \emph{query} \\
\texttt{GET /mastery}       & per-KC KT belief          & \emph{query} \\
\texttt{GET /activity}      & event log                 & \emph{query} \\
\texttt{GET /schedule}      & study blocks              & \emph{query} \\
\texttt{GET /unit\_content} & item catalog              & \emph{query} \\
\midrule
\multicolumn{3}{@{}l@{}}{\emph{LMS write}} \\
\texttt{POST /assign\_items}    & queue practice problems       & \texttt{assign\_items} \\
\texttt{POST /adjust\_level}    & difficulty $\pm$ for chosen KCs & \texttt{adjust\_level} \\
\texttt{POST /provide\_content} & push example / note / media   & \texttt{provide\_content} \\
\midrule
\multicolumn{3}{@{}l@{}}{\emph{Messaging}} \\
\texttt{GET /messages}   & poll learner replies    & \emph{query} \\
\texttt{POST /message}   & send a chat turn        & \texttt{send\_message} \\
(implicit)               & skip a turn             & \texttt{stay\_silent} \\
\bottomrule
\end{tabular}
\end{table}

\subsection{Simulated Learner}
\label{sec:student}

The simulated learner has two coupled components, namely a KT model that maintains a per-KC belief map, and an LLM that role-plays the student and produces free-response attempts.

\paragraph{Knowledge tracing.}
We train an Attentive Knowledge Tracing (AKT) model \cite{ghosh_2020_akt}, a lightweight model chosen for its stable and strong performance, on the XES3G5M elementary-mathematics interaction dataset \cite{liu_2023_xes3g5m,ozyurt_2024_kcqrl}, KC-level split (5{,}027 sequences truncated to 200 interactions each, 7{,}652 questions, 865 KCs, 5-fold learner-disjoint split). We train it under the 5-fold split, reaching a test AUC of $0.80$. At inference we export to CPU and maintain a belief $\mathbf{b}_{s,d} \in [0,1]^K$ over $K=831$ KCs (those that map to at least one usable item), updated at each observation of a question $q$, its knowledge concept $k$, and the graded response $y$ by
\begin{equation}
\mathbf{b}_{s,d+1} \;=\; f_{\theta}\bigl(\mathbf{b}_{s,d},\, q,\, k,\, y\bigr).
\label{eq:kt-update}
\end{equation}
The tutor reads $\mathbf{b}_{s,d}$ via \texttt{GET /mastery}.

\paragraph{Persona pool and initial-belief seeding.}
Every scenario is instantiated with one of $2{,}169$ persona seeds pooled from XES3G5M's training students. Each seed carries a 200-interaction warm-up history from that student's real practice log, and we precompute the induced initial belief $\mathbf{b}_{s,1}$ by replaying the warm-up through $f_\theta$ offline (one-time, roughly 30 minutes on GPU for the full $2{,}169 \times 831$ mastery matrix) and cache it. At scenario start we load the cached $\mathbf{b}_{s,1}$, and only Eq.~\ref{eq:kt-update} runs online. Because every persona seed starts from a distinct real practice history, no two runs share initial mastery.

\paragraph{LLM role-play.}
An LLM student generates $\hat{y} \sim p_{\text{LLM}}( \cdot \mid q, \mathbf{b}_{s,d}, c_{s,d} )$ on each probe, where $c_{s,d}$ is the tutor context. The system prompt is composed on the fly from three ingredients, namely the ability level (weak / average / strong), a grade band derived from the persona seed's XES3G5M metadata (roughly grades 5--8), and the weakest and strongest KCs extracted from $\mathbf{b}_{s,1}$ by thresholding. A pre-experiment holding items and tutor context fixed showed that the three ability levels produce significantly different pre-tutor probe accuracies (mean $0.19 / 0.34 / 0.53$ on a 200-item set), so the ability level is a real experimental knob. The same template is used verbatim across the four LLM student tiers of Section~\ref{sec:setup}, and only the underlying model changes. Graded probe outcomes are fed back into $f_\theta$, so KT and LLM stay coupled across $D=30$ virtual days.

\subsection{Evaluation Loop}
\label{sec:loop}

Algorithm~\ref{alg:eval} summarizes how the environment and the simulated learner compose on a single scenario. All state is written to a per-run trace log so that judge scoring in Section~\ref{sec:judge} can be re-run offline without re-executing the tutor.

\begin{algorithm}[t]
\caption{\emph{EduClaw-Bench} per-scenario evaluation loop}
\label{alg:eval}
\begin{algorithmic}[1]
\STATE $\mathbf{b}_1 \gets \textsc{InitKT}(\text{persona history})$
\FOR{$d = 1$ \TO $D$}
    \STATE $\text{pre}_d \gets \textsc{RunProbe}(\mathbf{b}_d, \text{LLM student})$
    \FOR{turn $t$ in day $d$}
        \STATE $u_{d,t} \gets \textsc{Tutor}(\text{context}_{d,t})$
        \STATE $\text{context}_{d,t+1} \gets \text{context}_{d,t} \cup \{ u_{d,t} \}$
    \ENDFOR
    \STATE $\text{post}_d \gets \textsc{RunProbe}(\mathbf{b}_d, \text{LLM student} \mid u_d)$
    \STATE $\mathbf{b}_{d+1} \gets f_\theta(\mathbf{b}_d, \text{probe outcomes})$
\ENDFOR
\STATE \textbf{return} $(\text{pre}_{1{:}D}, \text{post}_{1{:}D}, u_{1{:}D})$
\end{algorithmic}
\end{algorithm}

\begin{table*}[!t]
\centering
\scriptsize
\setlength{\tabcolsep}{3pt}
\caption{Untrained agent baseline. 10 agent adapters $\times$ 3 base-model tiers (the underlying LLM each adapter runs on, not an agent system), each row averaged over 4 students $\times$ 55 scenarios. The five axes are $\Delta$Solve ($\Delta$ Solve Rate, Axis I, deterministic), Resp (Axis II responsiveness, the fraction of student help-requests the tutor answers, rule-based), Help (Axis III, 29-item LearnLM rubric), Gagn\'e (Axis IVa, 9 events), and Rosen. (Axis IVb, Rosenshine, 10 principles). Help and the curriculum axes use a cross-family judge panel (gemini-3-flash, gpt-4.1-mini, kimi-k2.5), and Axis IV runs 1--5 with 1 meaning event absent. openclaw is the reference baseline in row 1. Within each column, \textbf{bold} marks the best value and \underline{underline} the second best. Answer-holding (0.05\% hand-over overall) and same-day responsiveness are reported in the text.}
\label{tab:baseline}
\begin{tabular}{l rrrrr rrrrr rrrrr}
\toprule
 & \multicolumn{5}{c}{Solar-pro3 (Frontier)} & \multicolumn{5}{c}{Codex-gpt5.5 (Frontier)} & \multicolumn{5}{c}{Qwen3-4B-Thinking (Small)} \\
\cmidrule(lr){2-6} \cmidrule(lr){7-11} \cmidrule(lr){12-16}
Adapter & $\Delta$Solve & Resp & Help & Gagn\'e & Rosen. & $\Delta$Solve & Resp & Help & Gagn\'e & Rosen. & $\Delta$Solve & Resp & Help & Gagn\'e & Rosen. \\
 & \%$\uparrow$ & \%$\uparrow$ & $\uparrow$ & $\uparrow$ & $\uparrow$ & \%$\uparrow$ & \%$\uparrow$ & $\uparrow$ & $\uparrow$ & $\uparrow$ & \%$\uparrow$ & \%$\uparrow$ & $\uparrow$ & $\uparrow$ & $\uparrow$ \\
\midrule
openclaw     & \textbf{+0.36} & $91.2$ & \textbf{6.11} & $1.46$ & $1.37$ & \underline{-0.27} & $90.5$ & \textbf{6.10} & $1.40$ & $1.34$ & $-0.28$ & $25.0$ & \textbf{6.53} & \underline{1.47} & \underline{1.42} \\
deeptutor    & $-0.65$ & $99.4$ & $4.55$ & $1.75$ & $1.50$ & $-1.25$ & $99.3$ & $4.58$ & $1.85$ & $1.56$ & $-0.98$ & \textbf{100.0} & $2.99$ & $1.26$ & $1.25$ \\
metaclaw     & $+0.02$ & \underline{99.7} & $4.23$ & $1.71$ & $1.50$ & $-0.90$ & \underline{99.7} & $4.26$ & \underline{1.90} & $1.59$ & \textbf{+0.64} & $96.5$ & \underline{4.73} & \textbf{1.74} & \textbf{1.62} \\
ironclaw     & $-0.94$ & $99.4$ & $4.54$ & \textbf{1.92} & \textbf{1.69} & $-0.74$ & \textbf{100.0} & $4.53$ & \textbf{1.93} & \textbf{1.69} & $-0.36$ & \textbf{100.0} & $3.04$ & $1.33$ & $1.31$ \\
picoclaw     & $-1.04$ & \textbf{99.8} & $3.76$ & \underline{1.85} & \underline{1.68} & $-0.77$ & \textbf{100.0} & $3.76$ & $1.85$ & \underline{1.67} & $-0.85$ & \textbf{100.0} & $2.97$ & $1.30$ & $1.28$ \\
zeroclaw     & \underline{+0.22} & $89.0$ & \underline{5.63} & $1.42$ & $1.40$ & \textbf{-0.08} & $89.5$ & \underline{5.66} & $1.44$ & $1.42$ & $-0.42$ & \textbf{100.0} & $3.12$ & $1.32$ & $1.30$ \\
hermesclaw   & $-0.75$ & $90.9$ & $5.58$ & $1.51$ & $1.45$ & $-1.08$ & $92.5$ & $5.62$ & $1.52$ & $1.46$ & \underline{-0.02} & \textbf{100.0} & $3.09$ & $1.32$ & $1.30$ \\
copaw        & $-0.56$ & $96.9$ & $5.02$ & $1.70$ & $1.55$ & \underline{-0.27} & $98.1$ & $4.97$ & $1.74$ & $1.58$ & $-0.46$ & \underline{99.5} & $2.87$ & $1.32$ & $1.31$ \\
nemoclaw     & $-1.24$ & $93.1$ & $5.13$ & $1.44$ & $1.39$ & $-0.78$ & $93.5$ & $5.06$ & $1.49$ & $1.43$ & $-0.63$ & \textbf{100.0} & $3.13$ & $1.27$ & $1.27$ \\
openclaw-rl  & $-0.63$ & $97.1$ & $4.84$ & $1.70$ & $1.55$ & $-0.49$ & $97.4$ & $4.86$ & $1.71$ & $1.56$ & $-0.85$ & \textbf{100.0} & $3.19$ & $1.24$ & $1.26$ \\
\bottomrule
\end{tabular}
\end{table*}

\subsection{Scoring Axes}
\label{sec:axes}

We score each completed run on five axes, namely three primary axes (I, II, III) plus two curriculum-design axes (IVa, IVb) reported alongside as a robustness check on pedagogical structure. Of these, only Axis I is inherited directly from prior pedagogical-RL work~\cite{lee_2026_pedagogicalrl_thinking,puerto_2025_pedagogicalrl}. We define the rest below and design the primary axes to be near-independent so that a tutor cannot game one at the expense of another.

Let a scenario $s$ have $D_s$ virtual days. On each day $d$ the tutor emits a sequence of turns $u_{s,d} = (u_{s,d,1}, \ldots, u_{s,d,T_{s,d}})$, and the student attempts pre-session and post-session probes. Denote the pre-session accuracy $\text{acc}^{\text{pre}}_{s,d}$ and post-session accuracy $\text{acc}^{\text{post}}_{s,d}$.

\subsubsection{Outcome and Interaction Axes (I, II, III).}

Axis I is the mean per-day learning gain, following the $\Delta$ Solve Rate formulation of \cite{lee_2026_pedagogicalrl_thinking} and given by
\begin{equation}
\Delta\text{SolveRate}(s) \;=\; \frac{1}{D_s} \sum_{d=1}^{D_s} \bigl( \text{acc}^{\text{post}}_{s,d} - \text{acc}^{\text{pre}}_{s,d} \bigr).
\label{eq:axis1}
\end{equation}

Axis II is Responsiveness, the fraction of student help-requests the tutor answers. Prior work instead uses a binary LeakJudge probed by adversarial students \cite{zhao2026leakage}, but that conflates whether the tutor engages a request with whether it hands over the answer, so we split the two, measuring engagement here as Responsiveness and disclosure below as answer-holding. Let $R_{s,d}$ be the set of student utterances on day $d$ that request help (a demand for the answer, a declaration of being stuck, or a request for an explanation), and let $\text{ans}(r) = 1$ if the tutor emits a message after request $r$. Then
\begin{equation}
\text{Responsiveness}(s) \;=\; \frac{\sum_{d} \sum_{r \in R_{s,d}} \text{ans}(r)}{\sum_{d} \lvert R_{s,d} \rvert}.
\label{eq:axis2}
\end{equation}
Because student utterances come from a fixed template set, request detection is a deterministic string match rather than a judge call, so Axis II is exact and carries no judge variance. We report same-day responsiveness (whether the tutor answers on the day of the request rather than later) as the discriminating variant. As an auxiliary safety signal we additionally score answer-holding on the days that contain a help-request, using a majority-of-three judge that labels each such day as $0$ (the tutor handed over the answer or a decisive step), $1$ (partial), or $2$ (held), and we report the rate of hand-over ($\text{score} \le 1$) as a leakage proxy.

Axis III is the per-day helpfulness averaged across days, given by
\begin{equation}
\text{Help}(s) \;=\; \frac{1}{D_s} \sum_{d=1}^{D_s} \frac{1}{3} \sum_{j=1}^{3} J^{(j)}_{\text{help}}(u_{s,d}),
\label{eq:axis3}
\end{equation}
where $J^{(j)}_{\text{help}}$ is judge $j$'s helpfulness score for the day, the mean of its 29 LearnLM rubric item scores in $[1, 10]$ \cite{learnlm_2025_v2}.

\subsubsection{Curriculum-Design Axes (IVa, IVb).}

We additionally report two curriculum-design axes grounded in the work of two foundational instructional-design theorists. Axis IVa scores each day (from $d=2$) on Gagn\'e's Nine Events of Instruction~\cite{gagne_1985_conditions}, and Axis IVb scores each day on Rosenshine's Ten Principles of Instruction~\cite{rosenshine_2012_principles}, both being LLM-judge rubrics on a $[1,5]$ scale where $1$ denotes ``event absent.'' Writing $r \in \{\text{IVa}, \text{IVb}\}$ for the two rubrics, each axis is the day-averaged, panel-averaged rubric score,
\begin{equation}
\text{Curric}_r(s) \;=\; \frac{1}{D_s - 1} \sum_{d=2}^{D_s} \frac{1}{3} \sum_{j=1}^{3} J^{(j)}_r\!\left(u_{s,d}, c_{s,d}\right),
\label{eq:axis4}
\end{equation}
where $J^{(j)}_r$ is judge $j$'s score for rubric $r$ (the mean of its nine or ten item scores in $[1,5]$) and $c_{s,d}$ is the day-context summary. Each day's judge prompt includes a compact summary of day $d-1$ (for Gagn\'e's event 3 and Rosenshine's principle 1, both of which require day-to-day continuity) and a rolling summary of the last seven days (for Rosenshine's principle 10, weekly review). We report both rubrics because they were designed for different audiences (Gagn\'e for instructional designers, Rosenshine as a research-based teaching guideline) and see high agreement in practice (Spearman $\rho > 0.97$), so treating both as columns lets a reader check that the curriculum signal is not an artifact of one rubric.

\subsection{LLM-as-Judge Protocol}
\label{sec:judge}

The Helpfulness, answer-holding, and two curriculum-rubric judges are all LLM judges, while Axis II needs none. Rather than trust a single model, we score every judged axis with a cross-family panel of three judges (gemini-3-flash, gpt-4.1-mini, and kimi-k2.5), each run with reasoning disabled at temperature $0.3$, taking the panel mean for Helpfulness and the curriculum rubrics and the panel majority for answer-holding. Three families guard against any one family's idiosyncratic scoring and keep a family from grading its own outputs when it appears as both student and judge, a failure an earlier single-judge configuration showed when its day-level scores were uncorrelated with the consensus. To calibrate the panel, two domain experts with a teaching or instructional-design background score stratified $40$-item samples per axis spanning the observed range, in Korean and blind to the panel. We report panel--expert agreement against the two experts' mutual agreement, the ceiling an automated judge can reach~\cite{zheng_2023_mtbench}, in Section~\ref{sec:results-human}.

%% file: sections/05_setup.tex

\section{Experimental Setup}
\label{sec:setup}

We evaluate 10 agent adapters, each an autonomous tutor harness, on three interchangeable base-model tiers that supply the underlying language model and bracket the design space, namely two closed-API frontier models (Solar-pro3 from Upstage and Codex-gpt5.5) and one small locally trainable open-weights model (Qwen3-4B-Thinking-2507, served through vLLM at roughly 8\,GB BF16) that enables the LoRA study of Section~\ref{sec:results}. A tier names the base language model an adapter runs on, not an agent system, so Codex-gpt5.5 is a frontier language model, not the coding agent of a similar name. The reported adapters are one full agent framework (deeptutor), a metaclaw skill-proxy, and eight Claw-family gateway adapters (openclaw, ironclaw, picoclaw, zeroclaw, hermesclaw, copaw, nemoclaw, openclaw-rl), with openclaw as the reference baseline and all non-in-process adapters isolated in Docker. Any tutor implementing the LMS interface plugs in the same way, including general-purpose coding agents, so the reported set spans controlled levels of scaffolding sophistication on one shared interface, from gateway adapters through a skill-proxy to the non-Claw deeptutor framework~\cite{deeptutor_2026}, isolating agent design from interface differences as in fixed-environment agentic benchmarks~\cite{yao_2024_taubench}. Two naive baselines and our under-development educlaw layer were run but excluded from the leaderboard as they are not agent frameworks. For each tier, four LLM students of mixed provenance (solar-mini, llama-3.1-8b-instruct, qwen-2.5-7b-instruct, gemma-3-4b-it) reduce single-family response bias, each seeing identical scenarios.

Every run uses seed 42 and fixes the student ability level at ``average,'' while the 55 scenarios draw distinct persona seeds, so initial mastery still varies across scenarios, with multi-seed confidence intervals left to future work. An internal experiment confirmed that pairing XES3G5M seeds with our different student models yields measurably different probe performance, so the learner pool spans differing abilities even with the ability level held fixed. The baseline budget is $4$ students $\times 13$ adapters $\times 55$ scenarios per tier, and we observed $8{,}637$ completed runs across the three tiers, the excess over the nominal budget being reruns of transient API failures, of which $6{,}626$ belong to the ten reported adapters. The Small tier's LoRA trajectory study (Section~\ref{sec:results}) adds $550$ runs, for $9{,}187$ in total. Full adapter inventory (ports, wrappers, prompt sources) is in the appendix.

%% file: sections/06_results.tex

\section{Results}
\label{sec:results}

\subsection{Benchmark Results}
\label{sec:results-outcome}

\begin{figure*}[!t]
\centering
\includegraphics[width=\textwidth]{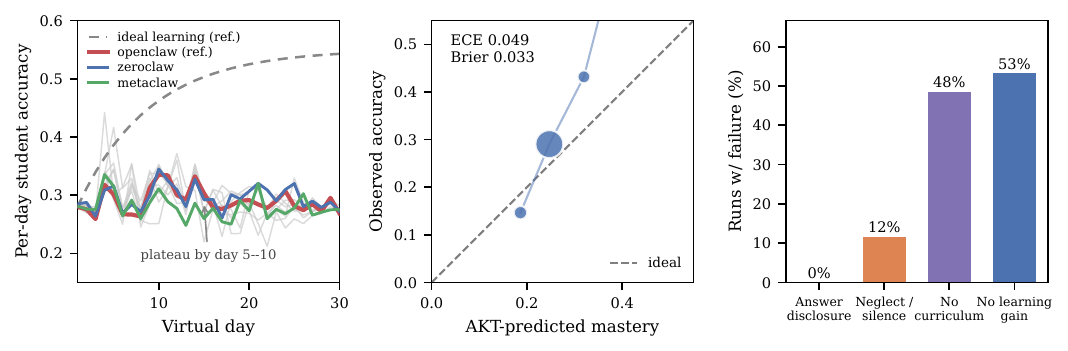}
\caption{\emph{Left} plots per-day student accuracy over 30 virtual days on the Codex tier, with three representative adapters in color, the other seven in gray, and a dashed ideal-learning reference. Every adapter plateaus by day five to ten far below the reference. \emph{Center} plots the simulator's AKT-predicted KC mastery against observed probe accuracy, and the points hug the diagonal (ECE $0.049$, Brier $0.033$ over $1.19$M attempts), so the simulated learner tracks a KT model fit on real students. \emph{Right} breaks tutor failures into four pedagogical modes and reports the share of runs in each, dominated by no-curriculum ($48.5\%$) and no-learning-gain ($53.3\%$) while answer disclosure is essentially absent ($\approx 0\%$).}
\label{fig:analysis}
\end{figure*}

Table~\ref{tab:baseline} reports the untrained baseline over 10 agent adapters and three tiers. No adapter leads Axis I on more than one tier (openclaw on Solar-pro3 $+0.36\%$, zeroclaw on Codex-gpt5.5 $-0.08\%$, metaclaw on Qwen3 $+0.64\%$), so base model and harness interact rather than contribute separably and a single-tier leaderboard mis-ranks the same systems~\cite{kapoor_2024_agentsmatter,zhu_2026_agenticeval}. Responsiveness spans $25$--$100\%$ and runs inversely to Helpfulness on the frontier tiers, where the always-answer adapters (picoclaw, deeptutor, ironclaw) score lowest on Help ($3.8$--$4.6$) and the withholding adapters (openclaw, zeroclaw) highest ($5.6$--$6.1$), while direct leakage stays negligible ($0.05\%$ hand-over). Over the full horizon the \emph{left} panel of Figure~\ref{fig:analysis} shows every agent plateaus within 5--10 days far below steady learning (openclaw $0.28\!\to\!0.29$).

The two curriculum-design axes (Gagn\'e, Rosenshine) ask whether an adapter treats the 30 days as a coherent sequence rather than independent sessions. ironclaw and metaclaw lead on the frontier tiers (Gagn\'e up to $1.93$), while on the Small tier only metaclaw keeps measurable structure (Gagn\'e $1.74$) and the rest fall to $1.24$--$1.47$, near the ``event absent'' floor. The two rubrics agree closely ($\rho > 0.97$), so we report Gagn\'e and keep Rosenshine as a robustness column. The Axis I leaders sit mid-pack, so learning gain does not imply pedagogical structure, and even the best scores ($1.93$ and $1.90$ against a $5.0$ ceiling) show that exposing the KT belief through the LMS does not force curriculum coherence.

Rejection-sampling fine-tuning (RFT) samples several candidate outputs per prompt, keeps the ones the reward scores highest, and fine-tunes on them. On the Small tier we apply it with LoRA (rank-16, 5 epochs) to metaclaw and openclaw-rl against a composite reward
\begin{equation}
R \;=\; 0.4\, R_{\text{I}} + 0.4\, R_{\text{II}} + 0.2\, R_{\text{III}},
\label{eq:composite-reward}
\end{equation}
over per-scenario normalized Axes I--III, with Table~\ref{tab:rft} reporting each adapter epoch-averaged ($n\!=\!275$). After training both adapters land near zero on Axis I yet diverge on helpfulness, as metaclaw-rft holds Help flat ($4.73 \to 4.76$) while openclaw-rl-rft collapses, its Help falling $3.19 \to 2.30$ (per-epoch $3.18 \to 2.50 \to 1.85$ by epoch~3, no recovery). Because neither loses responsiveness, this collapse is invisible to a leak- or refusal-based check and surfaces only on the pedagogy axes.

\begin{table}[t]
\centering
\scriptsize
\setlength{\tabcolsep}{3pt}
\caption{LoRA RFT on the Small tier (Qwen3-4B-Thinking), epoch-averaged ($n\!=\!275$), against the untrained Small tier baseline of the same two adapters from Table~\ref{tab:baseline}. Same five axes and judge panel.}
\label{tab:rft}
\begin{tabular}{l rrrrr}
\toprule
Adapter & $\Delta$Solve & Resp & Help & Gagn\'e & Rosen. \\
 & \%$\uparrow$ & \%$\uparrow$ & $\uparrow$ & $\uparrow$ & $\uparrow$ \\
\midrule
\multicolumn{6}{l}{\emph{Untrained (Small tier baseline)}} \\
metaclaw         & \textbf{+0.64} & \underline{96.5} & \underline{4.73} & \underline{1.74} & \underline{1.62} \\
openclaw-rl      & $-0.85$        & \textbf{100.0}  & $3.19$           & 1.24             & 1.26             \\
\midrule
\multicolumn{6}{l}{\emph{Trained (LoRA RFT, epoch-averaged)}} \\
metaclaw-rft     & \underline{+0.19} & $95.8$       & \textbf{4.76}    & \textbf{1.79}    & \textbf{1.63}    \\
openclaw-rl-rft  & $+0.10$        & \textbf{100.0}  & $2.30$           & 1.22             & 1.22             \\
\bottomrule
\end{tabular}
\end{table}

\subsection{Benchmark Reliability}
\label{sec:results-human}
Beyond the calibration of the simulated learner (ECE $0.049$, Section~\ref{sec:analysis}), we check the reliability of the judged measures from two directions, agreement with human experts and transfer to a real classroom.

\paragraph{Human validation.}
Two domain experts scored a stratified sample of about 40 judged days per axis blind to the panel (Section~\ref{sec:judge}), each giving one holistic score per day as the panel does. Table~\ref{tab:human} reports panel--expert Spearman agreement against the experts' mutual agreement, the ceiling an automated judge can reach~\cite{zheng_2023_mtbench}. The panel agrees with the experts on every judged axis (permutation $p<0.05$), reaching the expert ceiling on answer-holding and tracking Helpfulness and the two curriculum axes as a positive but more modest proxy.

\begin{table}[t]
\centering
\footnotesize
\setlength{\tabcolsep}{5pt}
\caption{Human validation of the judge panel (two domain experts, about 40 items per axis, blind to the panel). Panel--expert against expert--expert Spearman $\rho$, with a permutation $p$ for the panel--expert agreement.}
\label{tab:human}
\begin{tabular}{l c c c}
\toprule
Axis & Panel--exp.\ $\rho$ & Exp.--exp.\ $\rho$ & $p$ \\
\midrule
Helpfulness    & $0.33$ & $0.91$ & $0.04$ \\
Gagn\'e        & $0.43$ & $0.64$ & $0.008$ \\
Rosenshine     & $0.49$ & $0.90$ & $0.001$ \\
Answer-holding & $0.82$ & $0.85$ & ${<}0.001$ \\
\bottomrule
\end{tabular}
\end{table}

\paragraph{Field test.}
The second check asks whether our Helpfulness rubric measures the same way on real transcripts. We built a lightweight LMS that about 500 K-12 students used four times over two weeks, tutored by a stage-one regex claw on solar-mini that lies outside our simulation grid. Scoring 150 of these sessions with the same rubric gives a field Helpfulness of $6.02$ (sd $1.21$), inside the simulator's range for that axis ($6.19$, sd $0.45$) with no distinguishable difference (Mann-Whitney $p = 0.68$), and direct answer leakage matches at $0.05\%$ in both settings. The rubric therefore transfers to real transcripts, though since the field tutor is a different system measured in part by its own leak detector, this validates the instrument rather than any single adapter's score.

%% file: sections/07_analysis.tex

\section{Analysis and Ablations}
\label{sec:analysis}

\subsection{Can the measurement be trusted?}
\label{sec:analysis-defense}

The \emph{center} panel of Figure~\ref{fig:analysis} is the load-bearing check, since it asks whether the simulated learner behaves like a real student. Our KT-grounded student answers probes from an AKT model trained on real XES3G5M learners, and binning all $176{,}187$ (run, day) pairs by predicted mastery against observed accuracy gives an Expected Calibration Error of $0.049$ and a Brier score of $0.033$ over $1.19$M item attempts, so its knowledge state tracks the real-student KT model rather than drifting. The three primary axes are also largely independent, with pairwise Spearman correlations near zero on openclaw over the $n=30$ scenarios that contain a help-request ($\rho(\text{I},\text{II})\approx 0.01$, $\rho(\text{I},\text{III})\approx -0.12$, $\rho(\text{II},\text{III})\approx -0.13$), so a scalar composite would discard signal. Two robustness checks round this out, as the three-judge Helpfulness panel is internally consistent (mean pairwise Spearman $\rho=0.89$ across gemini-3-flash, gpt-4.1-mini, and kimi-k2.5) and the composite ranking is stable to leave-one-scenario-out resampling (Kendall $\tau$ mean $0.991$, min $0.911$ over 55 drops). Axes I and II carry no judge noise at all, being deterministic and rule-based.

\subsection{What does the benchmark reveal?}
\label{sec:analysis-utility}

The \emph{right} panel of Figure~\ref{fig:analysis} breaks tutor failures into four pedagogical modes. Answer disclosure is essentially absent ($\approx 0\%$, so leakage is not the operative failure) and neglect affects $11.7\%$ of runs, but the dominant failures are structural, with $48.5\%$ of runs showing no curriculum (Gagn\'e below $1.5$) and $53.3\%$ producing no learning gain ($\Delta$ Solve $\le 0$), and metaclaw is the only adapter to mostly escape the no-curriculum mode ($10\%$ of runs) while gateway adapters such as zeroclaw hit that mode on $82\%$ of runs. This structural failure is exactly what the long horizon exposes, since the per-day trajectories in the \emph{left} panel plateau by day five to ten far below steady learning, so almost no model-and-harness combination converts the full 30 days into sustained gain. Reliability tells the same story through pass$^k$, the fraction of scenarios where all $k$ of four independent student runs succeed. Following \citet{yao_2024_taubench}, every tier's pass$^1$ near $0.47$ collapses to pass$^4$ near $0.11$, so a tutor that helps the average student less than half the time almost never helps all four.

%% file: sections/09_discussion.tex

\section{Discussion and Conclusion}
\label{sec:discussion}

\emph{EduClaw-Bench} gives a vendor of a pedagogical LLM tutor an audit they can run before shipping to K-12 learners. Its three primary axes ($\Delta$ Solve Rate, Responsiveness, and the LearnLM Helpfulness rubric) plus two curriculum-design axes (Gagn\'e and Rosenshine) are near-independent in our data (Sections~\ref{sec:results} and~\ref{sec:analysis}), so no single reward is trivially game-able. Two findings follow that single-tier, single-session evaluation cannot reach. Tutoring quality is a property of the base model and the agent harness together rather than either alone, since adapter rankings reorder across tiers, so a vendor must evaluate the pairing rather than reuse a single-tier leaderboard. And almost no combination sustains good tutoring across the full horizon, since accuracy plateaus within days and curriculum structure rarely forms, so long-horizon competence needs a purpose-matched combination rather than a strong base model or a clever harness on its own. A fine-tuning study reinforces the point, as a low-weight helpfulness reward collapses one adapter family's pedagogy mid-training while staying invisible to leak- or refusal-based checks.

Two lines of evidence support trusting these measurements, each with a matching limit. Internally, the KT-grounded learner tracks real students to a calibration error of $0.049$ (Section~\ref{sec:analysis}), yet it remains a simulation rather than a full substitute for classroom learners. Externally, our Helpfulness rubric returns scores on the same scale for real K-12 transcripts as it does in simulation (Section~\ref{sec:results-human}), which shows the measurement is not an artifact of synthetic dialogue, though the field tutor is a different system covering one subject, so it validates the instrument, not a specific adapter's rank. We release all code, scenarios, per-run scores, and LoRA checkpoints, and recommend the benchmark as a pre-deployment gate rather than a substitute for supervised in-classroom evaluation, leaving a same-system field deployment to future work.

%% file: sections/10_appendix.tex

\section*{Appendix}
\appendix

\section{Scenario Inventory}
\label{app:scenarios}
The 55 scenarios are the Cartesian product of the 11 learner personalities and 5 study schedules in Table~\ref{tab:scenarios}, each run for 30 virtual days.

\begin{table}[h]
\centering
\footnotesize
\setlength{\tabcolsep}{3pt}
\caption{The 55 scenarios pair 11 learner personalities with 5 study schedules.}
\label{tab:scenarios}
\begin{tabular}{@{}ll@{}}
\toprule
Personality (11) & Schedule (5) \\
\midrule
confused productive     & after-school \\
disengaged              & homework-only \\
frustrated spiral       & school-integrated \\
gamer                   & weekend-warrior \\
help seeker             & self-directed \\
low SRL                 & \\
mastery approach        & \\
performance avoidance   & \\
silent struggler        & \\
wheel spinner           & \\
adversarial student     & \\
\bottomrule
\end{tabular}
\end{table}

\section{LearnLM Rubric}
\label{app:rubric}
Axis III scores tutor turns against 29 items following the LearnLM pedagogy rubric \cite{learnlm_2025_v2}, grouped into six principles. Each item is rated on a $1$ to $10$ scale by every judge, the item scores are averaged within a principle, and the principle scores are averaged to give a per-day Helpfulness score (Section~\ref{sec:axes}).

\begin{description}[leftmargin=1em, itemsep=1pt, topsep=2pt]
\item[Manages cognitive load (9 items).] Responses are an appropriate length; information is broken into manageable chunks; responses are clear and easy to follow; irrelevant information is avoided; narratives, case studies, or analogies are used effectively; information is presented in an appropriate style and structure; explanations develop in a logical order; information is not repeated unnecessarily; the tutor does not contradict earlier turns.
\item[Inspires active learning (4).] The tutor provides opportunities for engagement; asks questions that prompt the student to think; does not give away answers too quickly; promotes active engagement with the material.
\item[Deepens metacognition (4).] The tutor guides the student to discover their own mistakes; provides clear, constructive feedback; acknowledges when a response is correct; communicates a clear plan or objective.
\item[Stimulates curiosity (3).] The tutor tries to stimulate interest and curiosity; responds effectively when the student is frustrated; delivers feedback in an encouraging way.
\item[Adapts to needs (5).] Explanations are appropriate for the student's level; the tutor adapts its approach when the student is stuck; adapts to the student's needs overall; proactively guides the conversation when appropriate; does not withhold information unproductively.
\item[Overall (4).] There are no inaccuracies in the tutor's statements; the tutor expresses uncertainty when appropriate; it does not refuse reasonable questions; and, overall, it is as good as a very good human tutor.
\end{description}

\section{Adapter Details}
\label{app:adapters}
Every adapter runs as an isolated Docker service behind the LMS interface. Table~\ref{tab:adapters} lists each adapter's class, host port, and image. The eight Claw-family gateway adapters share a single \texttt{claw-bridge} image and differ only by a \texttt{CLAW\_NAME} environment variable that selects the adapter-specific prompt profile, while metaclaw and deeptutor use their own images. The base LLM is injected per tier through the \texttt{TUTOR\_MODEL} environment variable (Solar-pro3 or Codex-gpt5.5 on the frontier tiers, Qwen3-4B-Thinking on the Small tier), so the same adapter code runs unchanged across all three tiers. The per-adapter prompt profiles and wrapper entry points are included in the code release.

\begin{table}[t]
\centering
\small
\setlength{\tabcolsep}{5pt}
\caption{The 10 adapters, their class, Docker host port, and image. The eight gateway adapters share the \texttt{claw-bridge} image.}
\label{tab:adapters}
\begin{tabular}{@{}llrl@{}}
\toprule
Adapter & Class & Port & Image \\
\midrule
deeptutor    & framework    & 8002  & deeptutor \\
metaclaw     & skill-proxy  & 30000 & metaclaw \\
openclaw     & gateway      & 8010  & claw-bridge \\
ironclaw     & gateway      & 8011  & claw-bridge \\
picoclaw     & gateway      & 8012  & claw-bridge \\
zeroclaw     & gateway      & 8013  & claw-bridge \\
hermesclaw   & gateway      & 8014  & claw-bridge \\
copaw        & gateway      & 8015  & claw-bridge \\
nemoclaw     & gateway      & 8016  & claw-bridge \\
openclaw-rl  & gateway      & 8017  & claw-bridge \\
\bottomrule
\end{tabular}
\end{table}

\section{Additional Analyses}
\label{app:analyses}
These analyses extend Section~\ref{sec:analysis} and are not reported in the main paper. Unless stated otherwise they use the Codex tier, whose four students give the widest coverage.

\paragraph{Action profiles.}
Each run's trace records every tutor action, so we can summarize an adapter by how often it takes each of the five observable action kinds. Table~\ref{tab:action} reports the mean per-run frequency of each kind on the Codex tier, averaged over the four students and 55 scenarios. Three patterns stand out. First, the \texttt{adjust\_level} endpoint, which changes item difficulty for chosen knowledge concepts, is unused by all ten adapters (frequency exactly $0$), so the difficulty-adaptation affordance the environment exposes is left on the table by every harness we tested. Second, \texttt{provide\_content} spans two orders of magnitude, from metaclaw ($138.7$ per run) and ironclaw ($98.3$) down to zeroclaw ($11.3$) and openclaw ($17.3$), which is the clearest signature separating the content-pushing skill-proxy and framework adapters from the lightweight gateway adapters. Third, the gateway adapters that score high on Helpfulness (openclaw, zeroclaw, hermesclaw) spend a large share of their turns in \texttt{stay\_silent} ($22$ to $24$ per run), consistent with the withholding behavior that the Responsiveness and Helpfulness trade-off in Section~\ref{sec:results} attributes to them.

\begin{table}[t]
\centering
\scriptsize
\setlength{\tabcolsep}{3.5pt}
\caption{Mean per-run frequency of each tutor action kind (Codex tier). The \texttt{adjust\_level} endpoint is unused by every adapter.}
\label{tab:action}
\begin{tabular}{@{}lrrrrr@{}}
\toprule
Adapter & Msg & Silent & Assign & Adjust & Content \\
\midrule
openclaw     & 24.8 & 23.0 & 5.7  & 0 & 17.3 \\
deeptutor    & 29.1 & 16.3 & 20.9 & 0 & 61.2 \\
metaclaw     & 36.0 & 11.7 & 23.1 & 0 & 138.7 \\
ironclaw     & 46.3 & 0.3  & 32.5 & 0 & 98.3 \\
picoclaw     & 44.3 & 0.0  & 29.1 & 0 & 84.1 \\
zeroclaw     & 23.4 & 21.9 & 3.6  & 0 & 11.3 \\
hermesclaw   & 24.9 & 23.9 & 5.9  & 0 & 17.7 \\
copaw        & 34.4 & 12.7 & 17.6 & 0 & 54.3 \\
nemoclaw     & 29.3 & 20.0 & 10.2 & 0 & 31.5 \\
openclaw-rl  & 34.8 & 11.5 & 18.5 & 0 & 56.8 \\
\bottomrule
\end{tabular}
\end{table}

\paragraph{Responsiveness detail.}
Overall Responsiveness counts a help-request as answered if the tutor ever replies to it, whereas the same-day variant (Section~\ref{sec:axes}) counts it only if the reply lands on the day of the request. The gap between the two is a direct measure of how long an adapter makes a student wait. Table~\ref{tab:respdetail} shows that the overall rate collapses most adapters together near the top (eight of ten answer at least $90\%$ of requests eventually), so it discriminates poorly. The same-day rate pulls them apart. The withholding adapters that score highest on Helpfulness answer very few requests on the day they are made (zeroclaw $20.0\%$, hermesclaw $20.9\%$, openclaw $28.8\%$, a $60$ to $70$ point drop from their overall rate), while the always-answer adapters stay near $100\%$ on both (picoclaw $100.0\%$, ironclaw $99.8\%$). This is why we report same-day Responsiveness as the discriminating variant, and it cleanly separates the withholding gateway adapters from the always-answer ones.

\begin{table}[t]
\centering
\small
\setlength{\tabcolsep}{5pt}
\caption{Overall versus same-day Responsiveness (Codex tier).}
\label{tab:respdetail}
\begin{tabular}{@{}lrr@{}}
\toprule
Adapter & Overall & Same-day \\
\midrule
openclaw     & $90.5\%$  & $28.8\%$ \\
deeptutor    & $99.3\%$  & $94.8\%$ \\
metaclaw     & $99.7\%$  & $96.4\%$ \\
ironclaw     & $100.0\%$ & $99.8\%$ \\
picoclaw     & $100.0\%$ & $100.0\%$ \\
zeroclaw     & $89.5\%$  & $20.0\%$ \\
hermesclaw   & $92.5\%$  & $20.9\%$ \\
copaw        & $98.1\%$  & $78.1\%$ \\
nemoclaw     & $93.5\%$  & $40.9\%$ \\
openclaw-rl  & $97.4\%$  & $75.9\%$ \\
\bottomrule
\end{tabular}
\end{table}

\paragraph{Cross-student ranking stability.}
A benchmark is only useful if it ranks adapters consistently regardless of which student model role-plays the learner. To test this we rank the ten adapters within each of the four student models separately and take the mean pairwise Kendall $\tau$ across the four resulting rankings, per tier (Table~\ref{tab:crossstudent}). Helpfulness rankings are stable across students, with mean $\tau$ from $0.80$ on Solar-pro3 to $0.86$ on Qwen3 and a worst single pair of $0.61$, so which student we use barely changes who looks like a good tutor. $\Delta$ Solve Rate rankings are not stable, with mean $\tau$ from $-0.09$ on Codex to $0.24$ on Solar and a worst pair of $-0.42$, meaning the learning-gain ordering can invert entirely when the student model changes. The instability is expected, since $\Delta$ Solve Rate is a small per-cell difference at single seed and $n\approx220$ (Section~\ref{sec:results}), where noise dominates the near-zero true effects. The practical consequence is that Helpfulness is a reliable ranking anchor while $\Delta$ Solve Rate should be treated as a monitoring signal rather than a gate, which is exactly the recommendation in Appendix~\ref{app:deployment}. We omit the deprecated leakage axis from this comparison.

\begin{table}[t]
\centering
\small
\setlength{\tabcolsep}{6pt}
\caption{Mean pairwise Kendall $\tau$ of adapter rankings across the four student models, by tier.}
\label{tab:crossstudent}
\begin{tabular}{@{}lrrr@{}}
\toprule
Axis & Solar & Codex & Qwen3 \\
\midrule
Helpfulness         & $0.80$ & $0.84$ & $0.86$ \\
$\Delta$ Solve Rate & $0.24$ & $-0.09$ & $0.21$ \\
\bottomrule
\end{tabular}
\end{table}

\paragraph{Silent-run rates.}
Responsiveness is a ratio of answered help-requests to total help-requests (Section~\ref{sec:axes}), so a run in which the student issues no help-request the tutor engages with has an empty denominator. We report such a run as a silent run rather than assigning it an undefined or zero Responsiveness, which would otherwise bias the axis. The silent-run rate is negligible on the frontier tiers (Solar-pro3 $5.0\%$, Codex-gpt5.5 $2.6\%$) but rises sharply on the Small tier ($28.7\%$), and within that tier it reaches $90.0\%$ when the learner is role-played by qwen-2.5-7b. The pattern reflects that the weaker Small-tier base models frequently stay silent instead of acting, so their Responsiveness figures rest on far fewer scored runs than the frontier tiers and should be read together with this exclusion rate.

\paragraph{Horizon truncation.}
A short benchmark would rank adapters on their first few days rather than the full relationship. To quantify what that costs, we rank the ten adapters by mean per-day accuracy using only the first $N$ days and compare that ranking to the full 30-day ranking with Kendall $\tau$ (Table~\ref{tab:truncation}). The first five days give a ranking uncorrelated with, or even inverted from, the 30-day ranking ($\tau$ from $-0.42$ on Solar-pro3 to $0.20$ on Qwen3), and the agreement only climbs toward one as more days are included. Because every adapter plateaus at a nearly tied per-day accuracy (the day-30 means span only $0.276$ to $0.291$ on Solar, Section~\ref{sec:results}), these early rankings are dominated by noise, so this is ranking instability under near-flat scores rather than a reversal of a strong effect. Either way, the full horizon is needed to order adapters reliably.

\begin{table}[t]
\centering
\small
\setlength{\tabcolsep}{6pt}
\caption{Kendall $\tau$ between the adapter ranking from the first $N$ days and the full 30-day ranking, by tier.}
\label{tab:truncation}
\begin{tabular}{@{}lrrr@{}}
\toprule
Truncation & Solar & Codex & Qwen3 \\
\midrule
first 5 days  & $-0.42$ & $0.16$ & $0.20$ \\
first 10 days & $0.02$  & $0.42$ & $0.82$ \\
first 15 days & $0.07$  & $0.56$ & $0.82$ \\
first 20 days & $0.42$  & $0.69$ & $0.82$ \\
\bottomrule
\end{tabular}
\end{table}

\paragraph{Judge panel selection.}
Every judged axis uses the three-family panel (gemini-3-flash, gpt-4.1-mini, kimi-k2.5), run with reasoning disabled at temperature $0.3$, with the panel mean for Helpfulness and the curriculum rubrics and the majority for answer-holding. We excluded several candidate judges, notably solar-pro3, whose day-level scores were uncorrelated with the cross-family consensus (pairwise $-0.10$ to $-0.26$, from digit misreads), along with others that saturated the scale or failed to parse. On a rule-filtered three-point-scale check the retained three-judge panel reached $95.7\%$ unanimity.

\section{Checkpoints}
\label{app:ckpts}
Table~\ref{tab:ckpts} lists the per-epoch LoRA checkpoint steps for the two RFT adapters, archived alongside the code release.

\begin{table}[h]
\centering
\small
\setlength{\tabcolsep}{8pt}
\caption{Per-epoch LoRA checkpoint step for each RFT adapter.}
\label{tab:ckpts}
\begin{tabular}{@{}lccccc@{}}
\toprule
Adapter & Ep.\ 1 & Ep.\ 2 & Ep.\ 3 & Ep.\ 4 & Ep.\ 5 \\
\midrule
metaclaw-rft    & 45 & 90  & 135 & 180 & 225 \\
openclaw-rl-rft & 62 & 124 & 186 & 248 & 310 \\
\bottomrule
\end{tabular}
\end{table}

\section{Model and Judge Hyperparameters}
\label{app:hparams}
Table~\ref{tab:hparams} lists the final settings for the KT model, the LoRA fine-tuning, and the judge panel. The AKT model is trained on a single NVIDIA RTX 3090 under Ubuntu with Python 3.11 and PyTorch 2.x, and exported to CPU for inference, while the LLM students are served through vLLM in BF16 at roughly $8$\,GB.

\begin{table}[t]
\centering
\footnotesize
\setlength{\tabcolsep}{4pt}
\caption{Final hyperparameters for the KT model, LoRA fine-tuning, and judge panel.}
\label{tab:hparams}
\begin{tabular}{@{}lll@{}}
\toprule
Component & Setting & Value \\
\midrule
AKT      & model dim / heads / layers  & $64$ / $4$ / $2$ \\
AKT      & max sequence length         & $200$ \\
AKT      & epochs / CV folds           & $10$ / $5$ \\
LoRA RFT & rank / epochs               & $16$ / $5$ \\
LoRA RFT & reward weights (I/II/III)   & $0.4$ / $0.4$ / $0.2$ \\
Judge    & temp / reasoning / trials   & $0.3$ / off / $1$ \\
Judge    & aggregation (help/holding)  & mean / majority \\
\bottomrule
\end{tabular}
\end{table}

\section{Field Study Setup}
\label{app:field-setup}
About 500 K-12 students across several partner classrooms interacted with a live tutor through a purpose-built web UI. The deployed tutor is a stage-one regex claw on solar-mini, a real product configuration that is not one of the adapter-and-tier combinations in our simulation grid, so the field study checks whether our Helpfulness measurement transfers to real classroom transcripts rather than reproducing a specific leaderboard entry. IRB documentation is available upon request (anonymized in this submission).

\section{Field Study Log Schema and Judge Re-application}
\label{app:field-data}
Field logs record tutor turn text, LMS write events, and session timestamps in the same JSONL schema as our simulator traces, so the three-judge Helpfulness rubric and the answer-holding measure can be re-applied offline. From the sessions with at least three help turns we scored a random sample of 150 with the same rubric, giving a field Helpfulness of $6.02$ (sd $1.21$) against $6.19$ in simulation (Section~\ref{sec:results-human}), and a direct answer-leakage rate of $0.05\%$ ($5$ of $10{,}405$ tutor turns) that mirrors the simulator's $0.05\%$. The field leakage figure comes from the deployment's own regex detector, so it is a convergent rather than an identical-instrument check. Table~\ref{tab:field} reports the Helpfulness comparison, whose difference is not significant (Mann-Whitney $U=39{,}759$, $p=0.68$, Cohen's $d=0.18$).

\begin{table}[t]
\centering
\small
\setlength{\tabcolsep}{5pt}
\caption{Field versus simulation Helpfulness under the same LearnLM rubric.}
\label{tab:field}
\begin{tabular}{@{}lrrr@{}}
\toprule
Setting & $n$ & Help & sd \\
\midrule
Field (live) & $150$ & $6.02$ & $1.21$ \\
Simulation (openclaw) & $542$ & $6.19$ & $0.45$ \\
\bottomrule
\end{tabular}
\end{table}

\section{Field Study Teacher Interview Protocol}
\label{app:field-teacher}
Teachers provided free-form written feedback on students' interactions after the study period. The protocol asks for (i) failure modes observed, (ii) whether the tutor scaffolded rather than answered, and (iii) any adverse effects observed. We record the protocol for reproducibility, and a systematic analysis of the responses is left to future work.

\section{Deployment Recommendations}
\label{app:deployment}
We turn the results of Section~\ref{sec:results} into three recommendations for a vendor shipping an LLM tutor to K-12 learners.

\paragraph{Pre-deployment safety gate thresholds.}
We propose the following as an initial gate a vendor can add to a continuous-integration (CI) pipeline. First, block release if the answer-holding hand-over rate rises above its clean baseline. Answer-holding (the Axis II auxiliary) labels each help-request day by whether the tutor supplied the answer or a decisive step, and in our data hand-over occurs on only $0.05\%$ of such days, so a well-behaved tutor clears this gate with wide margin while a checkpoint that begins disclosing answers is caught immediately. Second, if Helpfulness is below $5.5$, warn. Below this line, most rubric items score at or below ``partially helpful,'' and the LearnLM anchors describe pedagogically problematic behavior \cite{learnlm_2025_v2}. Third, treat $\Delta$ Solve Rate as a monitoring signal only, not as a gate, because at single-seed and $n\approx220$ we observed inter-adapter variance of $\pm 1\%$ within noise, and gating on that axis produces false positives. Table~\ref{tab:gate} summarizes the three gates.

\begin{table}[h]
\centering
\small
\setlength{\tabcolsep}{4pt}
\caption{Pre-deployment safety gate a vendor can add to a CI pipeline.}
\label{tab:gate}
\begin{tabular}{@{}cllc@{}}
\toprule
Gate & Metric & Threshold & Action \\
\midrule
1 & Hand-over rate & above baseline & block \\
2 & Helpfulness & $< 5.5$ & warn \\
3 & $\Delta$ Solve Rate & $\pm 1\%$ (noise) & monitor \\
\bottomrule
\end{tabular}
\end{table}

\paragraph{Early-stopping rule for RFT tuners.}
The openclaw-rl-rft trajectory shows that Helpfulness decreases even as the composite training reward continues to rise. Tuners should therefore evaluate held-out Helpfulness after every epoch and roll back to the previous checkpoint whenever it drops. Applied to our data, this rule would have stopped openclaw-rl at epoch 2 (the observed Helpfulness peak) rather than shipping epoch 5. \citet{puerto_2025_pedagogicalrl} anticipated the same pattern, and our contribution is a per-epoch quantification that fixes an actionable stopping rule.

\paragraph{Adapter-selection guide by tier.}
Because adapter rank is tier-dependent (Section~\ref{sec:results-outcome}), we recommend that vendors pick a starting adapter by tier from our leaderboard, namely openclaw on frontier-Solar (Axis I $+0.36\%$, Helpfulness 6.11), zeroclaw on frontier-Codex (Axis I $-0.08\%$ tied first, Helpfulness 5.66), and metaclaw on the Small tier (Axis I $+0.64\%$, Helpfulness 4.73). Because Responsiveness and Helpfulness trade off (Section~\ref{sec:results-outcome}), a vendor that prioritizes scaffolding over immediate answers should prefer the withholding adapters (openclaw, zeroclaw) over the always-answer adapters (picoclaw, deeptutor), which reach near-$100\%$ Responsiveness but score $1.5$ to $2$ points lower on Helpfulness. We encourage vendors to re-run this benchmark on their own base model rather than assuming these choices generalize.

\section{Reproduction Commands}
\label{app:repro}
The code release reproduces the full pipeline through a single \texttt{bench} command-line interface. After installing the package and bringing up the adapter services with Docker Compose (Appendix~\ref{app:adapters}), \texttt{python -m bench run} executes one scenario against a chosen adapter and tier and writes a per-run trace log, \texttt{python -m bench score} re-applies the three-judge panel to the traces offline, and \texttt{python -m bench report} aggregates the per-run scores into the leaderboard. Parallel driver scripts run the full grid of 10 adapters, 3 tiers, and 4 students, and the persona seeds are regenerated from the XES3G5M warm-up histories by a separate script.

\section{Data Licensing and PII Notes}
\label{app:legal}
XES3G5M and its English translation KCQRL are used under their respective open licenses. Item text is redistributed under the same license, and no student identifiers are propagated.